Reframing the Galaxy and Cluster 'Mass Discrepancy' Problem:
A Consequence of Virial Equilibrium and Other Energy Considerations

Jeffrey M. La Fortune
1081 N. Lake St. Neenah, WI 54956 forch2@gmail.com
10 July 2023

*Abstract*
Galaxy and galaxy clusters exhibit tight robust physical scaling relations between baryons and system dynamics. One such phenomenon is 'mass discrepancy' with two leading solution spaces occupied by LCDM and MOND. Here, we propose an alternative solution to this puzzling problem exclusively based on application of the scalar virial theorem. For these dynamically equilibrated systems, we demonstrate there is ample virially-induced kinetic energy available to modify bulk structure dynamics in apparent violation of Newtonian law. We propose the ubiquitous Baryonic Tully-Fisher Relation represents the 'preferred' dynamic configuration that best assures long-term survivability for these thermodynamic quasi-equilibrated systems. We compare total mass estimates guided by the empirical evidence to those obtained from NFW dark matter halo fits ranging from small dwarf galaxies to massive galaxy clusters.

*Introduction/Background*
Almost a century ago, Zwicky observed the Coma galaxy cluster and discovered a significant discrepancy between luminous matter content and total mass derived from velocity dispersion observations based on the virial theorem (Zwicky 1933). At the time, a solid conjecture was made that the discrepant dispersion was due to an unseen matter component dominating cluster dynamics. A similar 'discrepancy' was detected by Slipher in M31 with first hints of an extended rotation profile in the galaxy's central region with subsequent surveys extending the rotation curve to the edge of the stellar disk (Slipher 1914) Two decades later Babcock crudely estimated the mass of M31 via its rotation curve and found it did not conform to Newtonian expectations (Babcock 1939). Rubin confirmed this same phenomenology for the M31's gas disk and other spiral galaxies, greatly extending rotation curves with attendant increases in dynamic mass (Rubin 1971).

Almost a century later, Zwicky's unseen mass remains the consensus solution to the so-called 'mass discrepancy' problem. Over the past fifty-years, an entire cosmology (LCDM) has been constructed to accommodate this dark matter component. For the most part, the viability of the dark matter solution currently hinges on the physical detection of dark matter beyond its gravitational influence. Presently, a world-wide search is underway comprising over two dozen experimental programs. The sheer size of this effort indicates the importance and perhaps urgency to find this missing mass component to conclusively resolve the 'mass discrepancy' problem.

In the early-eighties, an alternative to the 'mass discrepancy' problem was first promoted and termed MOdified Newtonian Dynamics (MOND) (Milgrom 1983). This approach eschews dark matter for modified gravitational law specifically attributed to disk galaxies. MOND and its variants are by and large the leading contender versus LCDM. Part of MOND's attractiveness is that can explain the Mass Discrepancy-Acceleration Relation (MDAR), Baryonic Tully-Fisher Relation (BTFR), and the Radial Acceleration Relation (RAR) (McGaugh 2016). As these scaling relations cannot be discounted nor ignored, inroads have been made demonstrating LCDM cosmology can also produce galaxies and clusters obeying these fundamental baryon-system relations (Chan 2017) (Ludow 2017) (Paranjape 2021) (Tam 2022).

While MOND has recently published preliminary hydrodynamic cosmological simulations and viral theorem analysis founded in alternative gravitational law, LCDM cosmology is more mature and consensus leader describing the physics behind galaxy and cluster formation and dynamics (Wittenburg 2023) (Lopez-Corredoria 2022). With the advent of the James Webb Space Telescope, previously accepted LCDM 'bedrock' tenets are now being challenged suggesting strong tension between LCDM predictions and observations. In light of these most recent findings, it is appropriate to revisit the problem of 'mass discrepancy' as it relates to the virial theorem, the basis of Zwicky's original discovery. Here, we reconstruct Zwicky's virial analysis but now armed with precision astrometrical data and directed by empirical scaling relations not available a century ago.

The expression 'mass discrepancy' has roots in the dark matter camp as it implies disagreement between estimated system mass obtained from bulk structure kinematics and baryon mass obtained via independent means. Per convention, 'mass discrepancy' is generally expressed as the ratio $D=M_{Dyn}/M_{Bar}$ with its LCDM equivalent, baryonic fraction $f_b=M_{Bar}/(M_{DM}+M_{Bar})$ where $M_{DM}$ is the virial mass of the dark matter halo responsible for non-Newtonian dynamics. With no dark matter support, MOND expressions are based on radial accelerations $a_{Dyn}/a_{Bar}$ and $V_C^2/V_{Bar}^2$ noting that mass, acceleration, and velocity squared terms used interchangeably in the 'mass discrepancy' expression depending on the motivation. For the past half century, LCDM and MOND have dominated the literature as the two most viable solutions to the 'mass discrepancy' problem. One approach not extensively explored to date are the complementary energy constraints and conditions imposed by the virial theorem. Below, we characterize 'mass discrepancy' in terms of kinetic and potential energies present in these virialized systems and compare our 'energy-motivated' total mass estimates against those obtained from supporting dark matter and modified gravity theory.

*'Scaling' Model Description and Approach*
Structurally, our simplified 'scaling' model encloses all mass and associated energies within baryon detected outermost radii. For disk galaxies, these radii ($R_D$) reference the measured extent of the HI gas disk ($R_D$). Likewise for clusters, outermost radii ($r_{last}$) correspond the furthest distance from their centers exhibiting 'constant' velocity dispersion for hydrostatic intracluster gas component. In outer regions, baryon velocities follow $1/\sqrt{r}$ decline. These are global parameters well suited for the scalar virial theorem avoiding uncertainties and mass model dependent extrapolations employing yet unsubstantiated cosmological arguments.

Rather than radially extended massive halos or modified gravity that predict flat rotation well beyond any last measured point, our simple model applies a mathematically sharp transition at outermost radii that breaks into a conventional Keplerian decline. We treat mass, velocity, and radius data as ideal and utilize existing/standard physical equations in a straightforward manner. This proposal extends and caps earlier work for SPARC galaxies and HIFLUGCS clusters listed in the appendices (La Fortune 2021a) (La Fortune 2021b).

Specifically, our approach is based on Zwicky's original analysis performed for the Coma cluster found in Section 5 in Zwicky's 1933 paper. Rather than presupposing a huge unseen mass responsible for this physical discrepancy, we focus on equivalent energies based on galaxy and cluster properties and their empirical scaling relations. For this analysis, we retain the informative dimensional richness inherent in kinematic expressions irretrievably lost upon conversion to equivalent halo mass and degeneracy issues that result. We posit virial regulation is the fundamental link between baryons and the characteristic scaling relations mentioned above.

*Energy Definitions, Equations, and General Properties*

In the classical realm, every object in motion has a kinetic energy attributed to it. Any large ensemble of gravitationally bound objects in dynamic equilibrium is composed of two time-averaged energy components, kinetic energy (T) and potential energy (U). The virial theorem assures that the ratio of the two system energy components is $2T=-U$.[1] Bound objects comprising the majority of mass (i.e., galactic stellar disks and cluster hydrostatic gas) conform to the $2T \leq -U$ energy condition with this data used to construct the MDAR, RAR, and the BTFR. Objects with high kinetic-potential energy ratios are limited by escape velocity with kinetic and potential energies in balance $T=-U$. High kinetic energy $T>-U$ objects are beyond escape velocity and are naturally 'scrubbed' from the data by time-averaging, leading to a very robust and universal constraint that is a core tenet of our proposal.

The first virially regulated energy condition is "total energy" or $2T=-U$ with kinetic energy content calculated from $M_{Bar}$ and disk ($V_C$, $R_D$) or cluster gas ($\sigma_{last}$, $r_{last}$). This relation was originally presented in a previous article (La Fortune 2021b). To discriminate between energy terms in this work, we rebadge "total energy" as dynamic energy $E_{Dyn}=T=R_D V_C^4/2G$ equivalent to $M_{Dyn}$ in the conventional 'mass discrepancy' expression. Likewise, the denominator $M_{Bar}$ is translated to baryonic kinetic $E_{Bar}=M_{Bar}V_C^2/2$. With replacement of mass terms for energy, the 'mass discrepancy' ratio is $D=(R_D V_C^4/2G)/(M_{Bar}V_C^2/2)$. Conversions between mass and energy are obtained with $M_{Dyn}=R_D V_C^2/G$ and $E_{Vir}=GM_{Vir}^2/R_{Vir}$, or alternatively $M_{Vir}=(E_{Vir} R_{Vir}/G)^{1/2}$. As the physics dictate, there is an exact correspondence between the two 'mass discrepancy' expressions holding the key that links this phenomenology to virial processes.

The second energy condition $T=-U$ is related to system escape velocity. Per our interpretation of the $M_{Dyn}$-$M_{Bar}$ scaling relation, this constraint fixes the virial energy component $E_{Vir}$ all dynamically equilibrated systems at $E_{Vir}=12.1E_{Bar}$ or its mass equivalent $M_{Tot}=12.1M_{Bar}$. The notion of 'mass discrepancy' becomes less mysterious when expressed as the ratio of kinetic energy captured in bulk baryon motion to total available kinetic energy or $R_D V_C^2/12.1GM_{Bar}$. For example, galaxies and clusters at the cosmic baryon fraction $f_b=0.17$ ($D=5.9$) have half of available kinetic energy ($E_{Vir}$) manifested in accelerated baryons (as $E_{Dyn}$), consistent with virial energy condition $2T=-U$.

The third overarching constraint is that all systems must obey the $V_C^4$-$M_{Bar}$ (BTFR) scaling relation in agreement with the empirical phenomenon. We propose the BTFR may actually be a necessary and sufficient condition that best assures long-term stability and presence in the $M_{Dyn}$-$M_{Bar}$ scaling relation.

We contrast our proposed dynamic against LCDM noting that $E_{Vir}$ is equivalent to $M_{Tot}$. Dark matter halo mass enclosed within outermost radii is simply $M_{DM}=M_{Tot}-M_{Bar}$, where $M_{DM}$=halo $M_{200}$. Within outermost radii, a significant portion of halo mass is precisely fixed by tracer data. Any halo mass above the physical requirement must reside outside only to be determined by extrapolation driven by theoretical assumptions imposed on the specific halo model. In a later section, we explore some implications that arise between Navarro-Frenk-White (NFW) halo fits and the robust escape velocity constraint featured in this work (Navarro 1997).

---

[1] Although we assume complete virial equilibrium, each component may exist in its own state of relaxation, further complicating total mass estimates amongst investigations (Yuan 2023). Yuan concludes intra-cluster gas is in a more relaxed state than the member galaxy population which exhibit "systematically large dynamical parameters" (noting dynamic mass estimates will proportionally differ as well). It could be possible that all components are in similarly relaxed states with kinetic energy being the primary 'dynamical parameter' that varies between them.

Note that our approach is similar to MOND phenomenology having the same dynamical characteristics with identical fits achieved using individually determined characteristic acceleration scale values that may vary from constant $a_0$. As MOND, we employ firmly established (but several new) scaling relations to guide our interpretation away from modified gravitational law as a solution for the 'mass discrepancy' problem toward one founded on existing physics and thermodynamic principles. Understanding that baryons carry the kinetic energy component, at outermost radii velocity support vanishes in contrast to the continuing MOND support (at $V_C$) to 'indefinite radii.' Provided virial, LCDM, and MOND deliver indistinguishable results from identical data, it presents a unique opportunity to exercise Occam's razor in a straightforward manner. Appendix A presents the Mass Discrepancy-Acceleration Relation (MDAR) for our SPARC and HIFLUGCS data listed in Appendices B and C, respectively (Lelli 2016) (Tian 2021).

*The $M_{Dyn}$-$M_{Bar}$ Scaling Relation*
On the next page in Figure 1 we present the $M_{Dyn}$-$M_{Bar}$ scaling relation (lower plot-left axis) and three energy analogs (upper plot-right axis) for 81 SPARC disk galaxies (black points) and 29 HIFLUGCS galaxy clusters (open black circles). The regressions are obtained from the combined sample treated as a single population. As a process check, we have reproduced Chan's $M_{Dyn}$-$M_{Bar}$ relation (Figure 2) obtained from a larger sample of the SPARC and HIFLUGCS catalogs (Chan 2022). Our non-weighted combined power law fit (green dash) has log slope γ=0.98 and zero-point D=7.7. Fixing our data to log slope γ=1 which the trend suggests, provides a number weighted mean 'mass discrepancy' D=5.9 intermediate between Chan's D=4.1 and D=8.2 for galaxies and clusters separately. Armed with independent corroboration, we demonstrate the $M_{Dyn}$-$M_{Bar}$ scaling relation holds the key to solve the 'missing mass' mystery. Our approach is to retain full dimensional richness and information inherent in the energy expressions underpinning the $M_{Dyn}$-$M_{Bar}$ relation. This effectively offers a higher-level solution based on the virial theorem avoiding baryon degeneracy issues and uncertainty prevalent in dark matter halo fits.

*New Complementary Energy Scaling Relations*
The $E_{Dyn}$-$M_{Bar}$ scaling relation was first published as "Total Energy" or system kinetic energy content ($E_{Dyn}$ herein) for the combined SPARC and HIFLUGCS datasets – see Figure 6 (La Fortune 2021b). Main features of this relation include unbroken $E_{Dyn} \sim M_{Bar}^{3/2}$ proportionality and close but not exact $M_{Dyn}$-$E_{Dyn}$ correspondence between individuals. At time of publication, this energy relation lacked proper context which is now remedied in Figure 1 below by the addition of two complementary energy relations. Using terms and equations defined earlier, first is $E_{Bar}$ serving the role of $M_{Bar}$ in the conventional 'mass discrepancy' expression. We find the $E_{Bar}$-$M_{Bar}$ relation to be a tight unbroken power law parallel slope to $E_{Dyn}$ and references the Newtonian kinetic energy associated with baryons. The second is $E_{Vir}$ or virial system energy is obtained from escape velocity considerations. For each sample in the lower plot, there are corresponding energies in the upper plots with $M_{Bar} \leftrightarrow E_{Bar}$ (open blue circles), $M_{Dyn} \leftrightarrow E_{Dyn}$ (purple triangles), and $M_{Tot} \leftrightarrow E_{Vir}$ (red crosses).

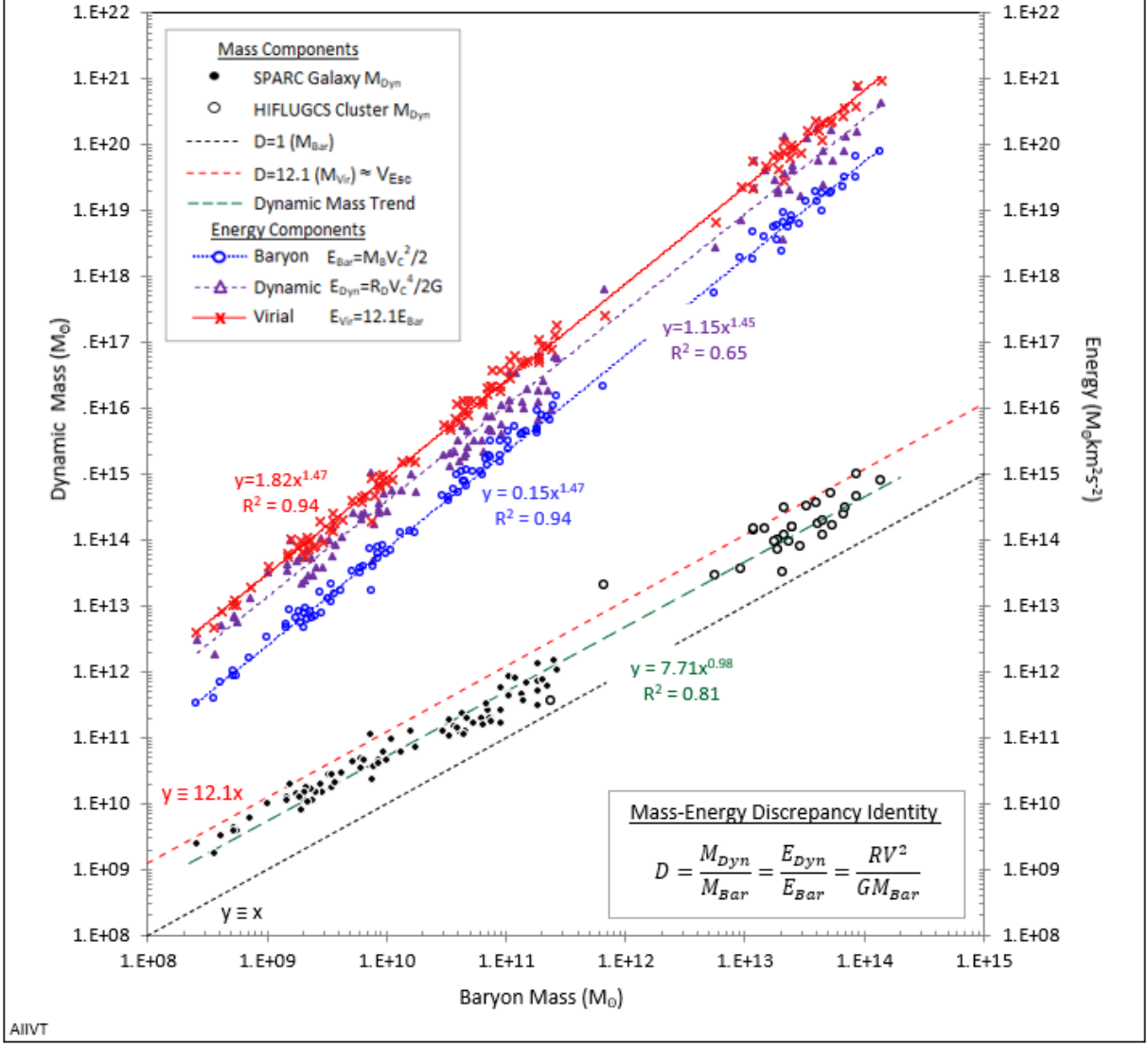

*Figure 1: Combined SPARC galaxy and HIFLUGCS cluster mass and energy scaling relations. Lower data-left axis – conventional 'mass discrepancy' relation (green dash) with regression fit. Upper data-right axis – system component energies with 1:1 (exact) correspondence with mass data. Energy expressions (identified in upper key) relate 'mass discrepancy' to observed properties (lower key) w/o dark matter or modified gravity. Data sources – (Lelli 2016) (Tian 2021)*

Inspecting Figure 1, the most significant and distinguishing feature in the $M_{Dyn}$-$M_{Bar}$ relation (lower plot) is the relatively narrow range of 'mass discrepancies' across all systems strictly limited from above by the strong upper cutoff $M_{Dyn}=12.1 M_{Bar}$ (red dash). This cutoff is universal and is a characteristic of the virial theorem ($GM_{Bar}$) with respective energy $E_{Vir}=12.1 E_{Bar}$ (solid red fit). Reference equations and regression fits are provided for combined galaxy and cluster data.

Evidence continues to mount this escape velocity cutoff in the MDARs of individual clusters. Li recently obtained two $M_{500}$ mass estimates for ten undisturbed HIFLUGCS clusters using different tracers to probe the gravitational potential (Li 2023). Li's "dynamic" method relies on member galaxy number density and velocity dispersion while the "hydro" method is determined X-ray surface brightness profiles for pressure supported intracluster gas (Table 1). We first focus on member galaxy radial acceleration profiles (Figure 6) noting the characteristic log slope γ=1 and maximum accelerations limited to below D=12.1 consistent with the $M_{Dyn}$-$M_{Bar}$ relation. In fact, Li's "dynamic" mean is D=11.2 in the range between D=5.7 and 16.7. Cluster member galaxies are loosely bound with the capacity to probe the true physical virial mass of their host clusters consistent with system escape velocity cutoff.

Turning to Li's "hydro" mass estimates, they can be substantially lower than "dynamic" estimates in agreement with the literature. With mean D=6.9 and less than half the variability, intra-cluster gas is unquestionably more tightly bound than member galaxies with this factor alone responsible for the discrepancy between methods. Although Li compared member galaxy acceleration profiles to the galactic RAR, we suggest future comparisons employ accelerations derived from the hydrostatic gas. This component dominants the energy-mass budget and shares considerably more characteristics with galactic disks than the cluster member galaxy cohort.[2]

We finish this section noting the exact correspondence between each 'mass discrepancy' value in the lower and upper plots shown in Figure 1. Rather than dark matter responsible for observed phenomena as implicitly implied in the $M_{Dyn}$-$M_{Bar}$, the physically operational relations are those expressed as energies. With knowledge that these energies are always present in virialized systems, we can explain 'mass discrepancy' simply as the amount ($E_{Dyn}$) of available kinetic energy ($E_{Vir}$) that is converted into bulk baryon motion in high energy density settings. We demonstrate with proper energy budget bookkeeping, a classical solution to the 'mass discrepancy' problem can be obtained. In the next section, we contrast total mass estimates for our SPARC and HIFLUGCS datasets derived from the energy relations and published NFW dark matter halo fits.

*Comparing Total Mass Estimates: The $M_{Dyn}$-$M_{Bar}$ Relation versus NFW Dark Matter Halos*
A cornerstone in our solution to the 'mass discrepancy' problem is that total virial energy $E_{Vir}$ (and equivalent total mass $M_{Tot}$) derived from the scaling relation is completely contingent on baryon content. This notion runs counter to LCDM cosmology with dark matter physics governing structure and dynamics and baryons serving a minor and semi-passive role in formation, evolution, and long-term survivability of these systems.

There are two issues fitting dark matter halos to reproduce baryon phenomena. The first is the coarse indirect method matching halo properties to the data. The NFW halo model contains no explicit parameter to independently fix $M_{Dyn}$ without impacting other primary halo parameters. The second issue is galaxy-halo degeneracy caused by the common unit of measure [$M_☉$] between baryons and dark matter. For example, even if $M_{Dyn}$ can be successfully reproduced by halo fits, the individual baryonic and dark matter contributions remain uncertain as does 'mass discrepancy.' We avoid these issues implementing the virial theorem with $D=R_D V_C^2/GM_{Bar}$ obtained directly from state properties.

---

[2] A similar issue has been noted for Milky Way Galaxy mass estimates. Independent of model, dynamic mass estimates derived from (tightly bound) stellar disk rotation curves (RCs) are typically a fraction of mass obtained from highly accelerated (loosely bound) hypervelocity star and dwarf satellite galaxy populations (La Fortune 2020) (La Fortune 2022). For the Galaxy this has resulted in total mass estimates differing between these two tracer cohorts by up to an order of magnitude (Jiao 2023).

At a minimum, every acceptable halo fit should be able to reproduce the observed level of velocity support for at least one $R_D$-$V_C$ pair of observations (without significant fine-tuning of observed baryon mass). Preferably, this radius should enclose as much halo mass as possible to reduce error in extrapolated halo mass estimates. Our only stipulation is that the modeled dark matter mass within this volume match the virial requirement assuring indistinguishable, exactly solvable results between solutions. A consequence in meeting this criterion is that galaxy and cluster dynamics automatically satisfy the BTFR.

For NFW halos in particular, any additional/excess halo mass residing beyond outermost radii (that does not contribute to the internal dynamic) may be considered a modelling artifact attributable to LCDM cosmological constraints. With empirical scaling relations as our guide, we expect the most "accurate" NFW dark matter halo mass estimates will fall along or just below the escape velocity cutoff relation.

We add one caveat regarding the definition of 'mass discrepancy.' The conventional expression is based on baryon energy $E_{Dyn}$ ($M_{Dyn}$ equivalent) which is not a formal halo parameter. As such an alternate expression is defined, baryon fraction $f_b = M_{Bar}/(M_{DM}+M_{Bar})$ where $M_{DM}$ is the $M_{200}$ (virial) halo mass and $M_{Dyn}=M_{DM}+M_{Bar}$. Although $D=M_{Dyn}/M_{Bar}$ is physically limited to a maximum near 12, it is not uncommon within LCDM for galaxies to sport massive halos with corresponding baryon fractions over 10x this value as it is based on the virial mass of the halo, not the baryon dynamical mass. We highlight this point with baryon fraction being a separate calculation lacking the $M_{Dyn}$ term in its expression.

In Figure 2, we compare total mass for NFW halo supported systems as $M_{Tot}=M_{200}+M_{Bar}$ versus $M_{Tot}=E_{Vir}$ relation as a function of baryon mass $M_{Bar}$. Here we employ published total mass for halo supported SPARC galaxies (black points) and HIFLUGCS clusters (open black circles) (Li 2020) (Reiprich 2002). As a process check, we include five recent $M_{200}$ cluster estimates (open blue circles) from Eckert for comparison to Reiprich values published over two decades earlier (Eckert 2022). The NFW halo fits are framed against the $M_{Dyn}$-$M_{Bar}$ relation with escape velocity cutoff D=12.1 (red dash) and Newtonian reference mass $M_{Bar}$ (black dot). These upper and lower limits bracket $M_{Dyn}$ values with the regression fit for the combined sample included (green dash).

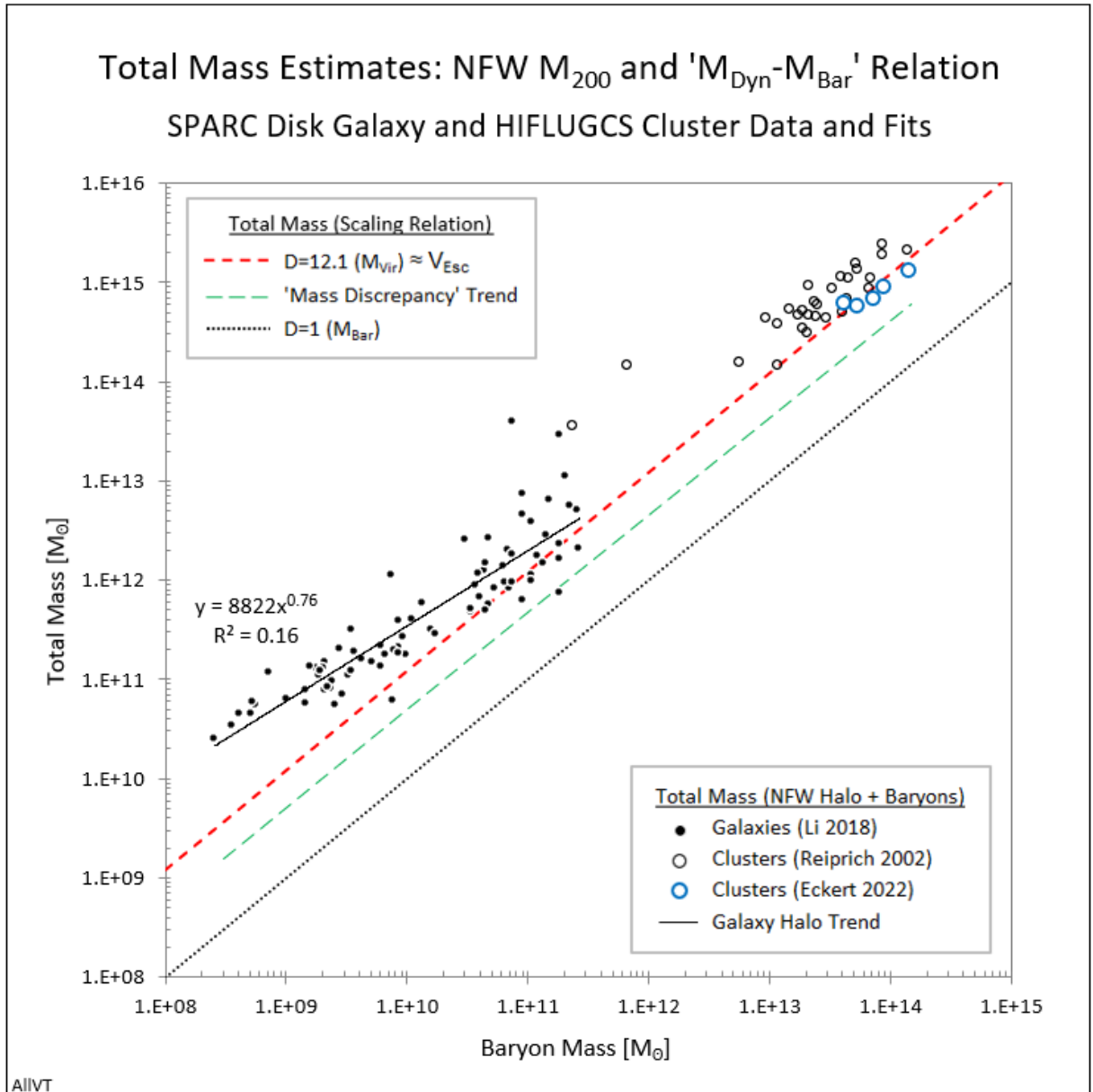

*Figure 2: Total Mass (Scaling Relation and NFW halo+baryons) as a function of Baryon Mass for SPARC galaxy and HIFLUGCS cluster datasets. Included are five recent cluster halo fits from Eckert for A85, A1795, A2029, A2142 and A3158. Cluster halos and high mass galaxies are consistent with escape velocity cutoff $M_{Vir}=12.1M_{Bar}$ and log slope γ≈1. Below $M_{Bar}$~$1x10^{11}$ $M_☉$, galaxy mass begins to diverge from the scaling relation with log slope γ=0.76 in agreement with Chan log slope γ=0.74. Sources - (Reiprich 2002) (Li 2020) (Eckert 2022) (Chan 2022)*

Visually inspecting the above figure, we see different halo mass functionalities depending on mass and structure. For galaxy clusters there is agreement between Reiprich's total mass obtained in halo fits and the $M_{Vir}$-$M_{Bar}$ scaling relation. Reiprich minimum cluster halo mass estimates run neatly along log slope γ=1, but tend to overestimate total mass by a factor of two compared to the escape velocity cutoff.

Reiprich results were obtained over a decade ago. In more recent work, five clusters in our HIFLUGCS sample have updated NFW $M_{200}$ halo mass estimates for comparison (Eckert 2022). Except for one stray, Eckert's latest cluster fits follow log slope γ=1 but in much better agreement in total mass with our expectation.

In Figure 2, the second trend concerns the NFW halo fits for SPARC galaxies with nearly all total masses exceeding the $M_{Dyn}$-$M_{Bar}$ relation. Below $M_{Bar}$~1x10$^{11}$ $M_\odot$, the deviation grows with dwarf galaxies containing ten-times more dark matter mass than established by escape velocity cutoff. If galaxy cluster dark matter halos fits are considered the gold standard, either galaxies behave much differently and/or the NFW model may not apply to disk galaxies. This is not an issue with virial regulation as it is universal, independent of mass, structure, or means of velocity support.

Galactic NFW-based total mass follows a shallower $M_{Tot}$-$M_{Bar}$ log slope γ=0.76 consistent with γ=0.74 recently obtained by Chan (Figure 1) based on $M_{500}$ halo mass (Chan 2022).[3] This behavior endemic and entirely dependent on theoretical cosmological considerations that neither virial processes or MOND are subject to. Consistent with the literature, the LCDM framework provides reasonable and accurate total mass estimates for massive relaxed galaxy clusters but remains challenged at the low end of the mass spectrum.

In a surprising result, it has been found that the HI gas mass in isolated disk galaxies (SPARC and LITTLE THINGS surveys) is a "direct tracer" of NFW $M_{200}$ halo mass with a clear 1:1 correlation (log slope γ≈1) independent of galactic mass or disk properties (Korsaga 2023). This latest investigation further bolsters confidence in the self-regulatory nature of the virial theorem as the source of 'mass discrepancy.' Although the authors contend this relation still needs to be understood, we see this as another facet of strict baryon virial regulation, unappreciated since Zwicky's first encounter with the Coma Cluster.

Physically consistent dark matter mass estimates such as those from Eckert have been touted as evidence for the physical existence of halos and by extension, the entire LCDM cosmological framework (Darragh-Ford 2023). However, we take a different tact and state that for amenable structure and dynamics (undisturbed galaxy clusters), halo models are only now finding consistency with firmly established scaling relations, the physical benchmarks.

*Summary*
The virial theorem offers a solution for the unresolved 'mass discrepancy' problem discovered a century ago. Our approach is advantaged over LCDM and MOND in that this proposed solution does not invoke unseen matter or violate Newtonian law. We argue 'mass discrepancy' is the secular baryon response (acceleration boost) sourced from inherent kinetic energy present within the virial volume. We also surmise the BTFR may simply reflect the principle of least action operating in these stably conserved maximally entropic systems.

*Acknowledgements*
We thank authors, arXiv, and professional journals for their willingness to publish research making it available for all that have an interest. This work is dedicated to the author's father and grandsons.

---

[3] We fail to reproduce Chan's $M_{500}$-$M_{Bar}$ galaxy cluster slope γ=0.75 with our $M_{200}$ values consistent with γ=1 (Chan 2022).

*Appendix A: The Mass Discrepancy-Acceleration Relation for SPARC Galaxy and HIFLUGCS Clusters*
Figure 3 below is the MDAR plot for our combined sample, identified separated (upper key). Regression fits and power-law equations are provided for galaxies (gray long dash) and galaxy clusters (gray dot-dash). The robust $M_{Dyn}$-$M_{Bar}$ scaling relation escape velocity cutoff D=12.1 (red dash) and D=5.9, the cosmic baryon fraction $f_b$=0.17 (green dash). The galactic RAR (solid gray) derived from SPARC galaxy data based on MOND universal acceleration constant $a_0$ (McGaugh 2016).

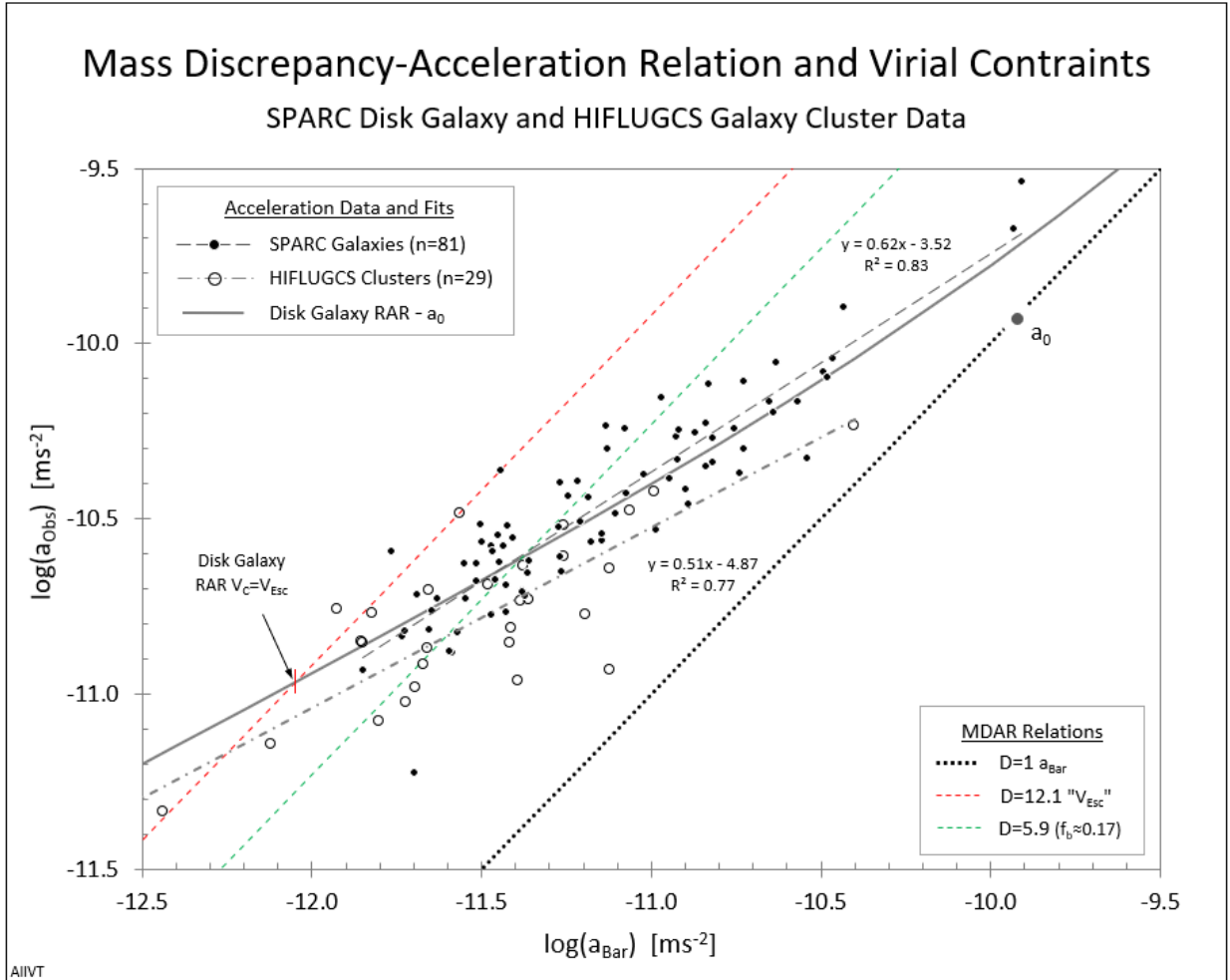

*Figure 3: Mass Discrepancy-Acceleration Relation (MDAR) for SPARC disk galaxies and HIFLUGCS galaxy clusters listed in Appendix B. Galaxy and cluster data is consistent with the BTFR with regressions demonstrating general RAR functionality. Our galaxy RAR distribution exhibits higher scatter than the MOND RAR obtained from the binned average of many full rotation curves. Sources – (Lelli 2016) (McGaugh 2016) (Tian 2021)*

Captured below are several important features in Figure 3, with comments:
- At outermost radii, clusters exhibit lower characteristic accelerations than their galactic counterparts contradicting many independent studies. The difference is a selection artifact with maximal cluster accelerations (>10x $a_0$) broadly limited to intermediate acceleration regions where the galactic RAR characterized by its smooth monotonic decline and constant MOND characteristic acceleration scale $a_0$. At lowest accelerations the RARs converge with comparable 'mass discrepancy' values near the cosmic mean.

- Outermost radii galaxy and cluster accelerations are widely distributed around the cosmic baryon fraction (D=5.9) with both regression fits broadly consistent with the galactic RAR. Only a few data points fall precisely on the RAR implying the vast bulk of cosmic systems will not be consistent with MOND predictions based on characteristic acceleration constant $a_0$. This variability in acceleration scales has been quantified and found to be dependent on structural and dynamic properties of individual systems.[4]
- Save for a few, minimum galaxy accelerations do not fall below the cluster "RAR" fit for the entire range investigated. If not a sampling artifact, this may be the physical lower acceleration limit for disk galaxies not observed for clusters.
- The abrupt truncation in the SPARC data at low accelerations has been previously noted but lacking explanation. Referring to Figure 3 above, we gain a physical understanding by tracing the RAR to the left ($V_C$ as a function of radius) until it intersects with D=12.1 (local $V_{Esc}$). Beyond terminal acceleration $a_{Bar} \approx 8 \times 10^{-13}$ ms$^{-2}$, stable disks are not assured.
- Cluster hydrostatic gas accelerations measured at outermost radii exhibit a weak power-law correlation with log slope γ=0.51, consistent with the galactic $a_{Dyn}$-$a_{Bar}$ slope predicted by MOND in the low acceleration regime and suggests similarity between disk and gas components. From a dynamical perspective, intracluster gas is preferred over member galaxies when comparing 'mass discrepancy' or RAR profiles between galaxies and clusters. We emphasize Li's work noting the member galaxy cohort better traces the virial (or total mass) properties of clusters than does intracluster gas. Member galaxy acceleration profiles better trace the cluster escape velocity envelope with 'asymptotic' log slope γ=1.

Despite the relatively high scatter in the MDAR acceleration space, robust scaling relations are encoded within all these systems, regardless of motivation. The empirical evidence leads us to consider the universal (and surprisingly strict) virial theorem to be fundamentally responsible for disk galaxy and galaxy cluster 'mass discrepancy' phenomena.

---

[4] For example, galaxy (and cluster) BTFRs are exactly solvable with $M_{Bar}=V_C^4/(\pi G^2 D \Sigma_{Dyn})$ where $\Sigma_{Dyn}=M_{Dyn}/\pi R_D^2$ and $M_{Dyn}=R_D V_C^2/G$ with all parameters fixed by observation. This equation treats the embedded acceleration term $a= \pi G^2 D \Sigma_{Dyn}$ as an output (La Fortune 2021a) (La Fortune 2021b)

*Appendix B: SPARC Disk Galaxy Data, Parameters, and Working Equations*

Spitzer Photometry and Accurate Rotation Curves (SPARC): Quality Flag=1 (Lelli, 2016)

| ID | Model Input | | | Mass and Energy Parameters | | | | | |
|---|---|---|---|---|---|---|---|---|---|
| Galaxy Name | Baryon Mass | Velocity Dispersion | Last Radius | Dynamic Mass | Dynamic Energy | Baryon Energy | 'Mass Discrepancy' | Virial Energy | Total Mass |
| n=81 | $M_{Bar}$ | $V_C$ | $R_D$ | $M_{Dyn}=R_D V_C^2/G$ | $E_{Dyn}=R_D V_C^4/2G$ | $E_{Bar}=M_{Bar}V_C^2/2$ | $D=M_{Dyn}/M_{Bar}$ $D=E_{Dyn}/E_{Bar}$ | $E_{Vir}=12.1 E_{Bar}$ | $M_{Vir}=12.1 M_{Bar}$ |
| [ID] | [$M_\odot$] | [kms$^{-1}$] | [kpc] | [$M_\odot$] | [$M_\odot$km$^2$s$^{-2}$] | [$M_\odot$km$^2$s$^{-2}$] | [Ratio] | [$M_\odot$km$^2$s$^{-2}$] | [$M_\odot$] |
| UGC02885 | 2.57E+11 | 289.5 | 74.2 | 1.45E+12 | 6.06E+16 | 1.08E+16 | 5.6 | 1.30E+17 | 3.11E+12 |
| UGC06614 | 9.12E+10 | 199.8 | 60.6 | 5.63E+11 | 1.12E+16 | 1.82E+15 | 6.2 | 2.20E+16 | 1.10E+12 |
| UGC09133 | 1.86E+11 | 226.8 | 60.4 | 7.22E+11 | 1.86E+16 | 4.79E+15 | 3.9 | 5.79E+16 | 2.25E+12 |
| ESO563-G021 | 1.86E+11 | 314.6 | 55.7 | 1.28E+12 | 6.35E+16 | 9.21E+15 | 6.9 | 1.11E+17 | 2.25E+12 |
| NGC6674 | 1.51E+11 | 241.3 | 50.0 | 6.77E+11 | 1.97E+16 | 4.41E+15 | 4.5 | 5.33E+16 | 1.83E+12 |
| NGC2841 | 1.07E+11 | 284.8 | 45.1 | 8.51E+11 | 3.45E+16 | 4.35E+15 | 7.9 | 5.26E+16 | 1.30E+12 |
| NGC0801 | 1.86E+11 | 220.1 | 45.0 | 5.07E+11 | 1.23E+16 | 4.51E+15 | 2.7 | 5.46E+16 | 2.25E+12 |
| NGC2998 | 1.07E+11 | 209.9 | 43.6 | 4.47E+11 | 9.84E+15 | 2.36E+15 | 4.2 | 2.86E+16 | 1.30E+12 |
| UGC11455 | 2.04E+11 | 269.4 | 43.4 | 7.33E+11 | 2.66E+16 | 7.41E+15 | 3.6 | 8.97E+16 | 2.47E+12 |
| NGC6195 | 2.24E+11 | 251.7 | 40.9 | 6.02E+11 | 1.91E+16 | 7.09E+15 | 2.7 | 8.58E+16 | 2.71E+12 |
| UGC02487 | 2.69E+11 | 332.0 | 40.2 | 1.03E+12 | 5.68E+16 | 1.48E+16 | 3.8 | 1.79E+17 | 3.26E+12 |
| NGC5985 | 1.20E+11 | 293.6 | 39.5 | 7.92E+11 | 3.41E+16 | 5.18E+15 | 6.6 | 6.27E+16 | 1.45E+12 |
| NGC3198 | 3.39E+10 | 150.1 | 35.7 | 1.87E+11 | 2.10E+15 | 3.82E+14 | 5.5 | 4.62E+15 | 4.10E+11 |
| NGC5055 | 9.12E+10 | 179.0 | 35.1 | 2.61E+11 | 4.19E+15 | 1.46E+15 | 2.9 | 1.77E+16 | 1.10E+12 |
| NGC1003 | 1.12E+10 | 109.8 | 33.3 | 9.34E+10 | 5.63E+14 | 6.76E+13 | 8.3 | 8.18E+14 | 1.36E+11 |
| NGC3992 | 1.35E+11 | 241.0 | 32.8 | 4.42E+11 | 1.28E+16 | 3.92E+15 | 3.3 | 4.74E+16 | 1.63E+12 |
| IC4202 | 1.07E+11 | 242.6 | 32.1 | 4.40E+11 | 1.29E+16 | 3.15E+15 | 4.1 | 3.82E+16 | 1.30E+12 |
| UGC00128 | 1.58E+10 | 129.3 | 31.3 | 1.22E+11 | 1.02E+15 | 1.32E+14 | 7.7 | 1.60E+15 | 1.92E+11 |
| NGC1090 | 4.79E+10 | 164.4 | 30.5 | 1.92E+11 | 2.59E+15 | 6.47E+14 | 4.0 | 7.83E+15 | 5.79E+11 |
| NGC5371 | 1.86E+11 | 209.5 | 30.0 | 3.07E+11 | 6.73E+15 | 4.09E+15 | 1.6 | 4.94E+16 | 2.25E+12 |
| NGC5033 | 7.08E+10 | 194.2 | 29.5 | 2.59E+11 | 4.88E+15 | 1.33E+15 | 3.7 | 1.62E+16 | 8.57E+11 |
| UGC03205 | 6.92E+10 | 219.6 | 28.6 | 3.21E+11 | 7.73E+15 | 1.67E+15 | 4.6 | 2.02E+16 | 8.37E+11 |
| NGC7331 | 1.41E+11 | 239.0 | 27.0 | 3.59E+11 | 1.02E+16 | 4.03E+15 | 2.5 | 4.88E+16 | 1.71E+12 |
| F571-8 | 7.41E+09 | 139.7 | 24.6 | 1.11E+11 | 1.09E+15 | 7.23E+13 | 15.0 | 8.75E+14 | 8.97E+10 |
| NGC4157 | 6.31E+10 | 184.7 | 24.1 | 1.91E+11 | 3.26E+15 | 1.08E+15 | 3.0 | 1.30E+16 | 7.63E+11 |
| UGC07125 | 7.59E+09 | 65.2 | 23.0 | 2.28E+10 | 4.84E+13 | 1.61E+13 | 3.0 | 1.95E+14 | 9.18E+10 |
| NGC4088 | 6.46E+10 | 171.7 | 22.3 | 1.53E+11 | 2.25E+15 | 9.52E+14 | 2.4 | 1.15E+16 | 7.81E+11 |
| UGC05005 | 6.17E+09 | 98.9 | 21.6 | 4.92E+10 | 2.40E+14 | 3.02E+13 | 8.0 | 3.65E+14 | 7.46E+10 |
| NGC6946 | 4.07E+10 | 158.9 | 21.3 | 1.25E+11 | 1.58E+15 | 5.14E+14 | 3.1 | 6.22E+15 | 4.93E+11 |
| NGC4559 | 1.74E+10 | 121.2 | 21.2 | 7.23E+10 | 5.31E+14 | 1.28E+14 | 4.2 | 1.54E+15 | 2.10E+11 |
| NGC3893 | 3.72E+10 | 174.0 | 20.8 | 1.47E+11 | 2.22E+15 | 5.62E+14 | 3.9 | 6.81E+15 | 4.50E+11 |
| UGC06786 | 4.37E+10 | 219.4 | 20.3 | 2.27E+11 | 5.47E+15 | 1.05E+15 | 5.2 | 1.27E+16 | 5.28E+11 |
| NGC3521 | 4.79E+10 | 213.7 | 18.9 | 2.00E+11 | 4.57E+15 | 1.09E+15 | 4.2 | 1.32E+16 | 5.79E+11 |
| UGC03546 | 5.37E+10 | 196.9 | 18.4 | 1.66E+11 | 3.21E+15 | 1.04E+15 | 3.1 | 1.26E+16 | 6.50E+11 |
| NGC0891 | 7.59E+10 | 216.1 | 18.2 | 1.97E+11 | 4.61E+15 | 1.77E+15 | 2.6 | 2.14E+16 | 9.18E+11 |
| NGC4100 | 3.39E+10 | 158.2 | 18.1 | 1.05E+11 | 1.32E+15 | 4.24E+14 | 3.1 | 5.13E+15 | 4.10E+11 |
| ESO079-G014 | 3.02E+10 | 175.0 | 17.7 | 1.26E+11 | 1.93E+15 | 4.62E+14 | 4.2 | 5.60E+15 | 3.65E+11 |
| NGC3953 | 7.41E+10 | 220.8 | 17.4 | 1.97E+11 | 4.80E+15 | 1.81E+15 | 2.7 | 2.19E+16 | 8.97E+11 |
| UGC06930 | 8.71E+09 | 107.2 | 16.8 | 4.48E+10 | 2.57E+14 | 5.00E+13 | 5.1 | 6.06E+14 | 1.05E+11 |

*Appendix B: SPARC Disk Galaxy Data, Parameters, and Working Equations – cont'd*

(cont'd)  Spitzer Photometry and Accurate Rotation Curves (SPARC): (Lelli, 2016)

| ID | Model Input | | | Mass and Energy Parameters | | | | | |
|---|---|---|---|---|---|---|---|---|---|
| Galaxy Name | Baryon Mass | Velocity Dispersion | Last Radius | Dynamic Mass | Dynamic Energy | Baryon Energy | 'Mass Discrepancy' | Virial Energy | Total Mass |
| n=81 | $M_{Bar}$ | $V_C$ | $R_D$ | $M_{Dyn}=R_D V_C^2/G$ | $E_{Dyn}=R_D V_C^4/2G$ | $E_{Bar}=M_{Bar}V_C^2/2$ | $D=M_{Dyn}/M_{Bar}$ $D=E_{Dyn}/E_{Bar}$ | $E_{Vir}=12.1 E_{Bar}$ | $M_{Vir}=12.1 M_{Bar}$ |
| [ID] | [$M_\odot$] | [kms$^{-1}$] | [kpc] | [$M_\odot$] | [$M_\odot$km$^2$s$^{-2}$] | [$M_\odot$km$^2$s$^{-2}$] | [Ratio] | [$M_\odot$km$^2$s$^{-2}$] | [$M_\odot$] |
| NGC4217 | 4.57E+10 | 181.3 | 16.7 | 1.28E+11 | 2.10E+15 | 7.51E+14 | 2.8 | 9.09E+15 | 5.53E+11 |
| NGC0100 | 4.27E+09 | 88.1 | 16.4 | 2.95E+10 | 1.15E+14 | 1.66E+13 | 6.9 | 2.00E+14 | 5.16E+10 |
| F574-1 | 7.94E+09 | 97.8 | 16.2 | 3.60E+10 | 1.72E+14 | 3.80E+13 | 4.5 | 4.60E+14 | 9.61E+10 |
| NGC4183 | 1.00E+10 | 110.6 | 16.1 | 4.57E+10 | 2.80E+14 | 6.12E+13 | 4.6 | 7.40E+14 | 1.21E+11 |
| UGC06983 | 6.61E+09 | 109.0 | 16.1 | 4.44E+10 | 2.64E+14 | 3.92E+13 | 6.7 | 4.75E+14 | 7.99E+10 |
| F583-1 | 3.31E+09 | 85.8 | 15.7 | 2.68E+10 | 9.86E+13 | 1.22E+13 | 8.1 | 1.47E+14 | 4.01E+10 |
| NGC2403 | 9.33E+09 | 131.2 | 15.1 | 6.05E+10 | 5.21E+14 | 8.03E+13 | 6.5 | 9.72E+14 | 1.13E+11 |
| F568-V1 | 5.25E+09 | 112.3 | 14.4 | 4.22E+10 | 2.66E+14 | 3.31E+13 | 8.0 | 4.00E+14 | 6.35E+10 |
| NGC3917 | 1.35E+10 | 135.9 | 14.1 | 6.05E+10 | 5.58E+14 | 1.25E+14 | 4.5 | 1.51E+15 | 1.63E+11 |
| NGC6503 | 8.71E+09 | 116.3 | 14.1 | 4.42E+10 | 2.99E+14 | 5.89E+13 | 5.1 | 7.13E+14 | 1.05E+11 |
| NGC2903 | 4.47E+10 | 184.6 | 13.8 | 1.09E+11 | 1.86E+15 | 7.61E+14 | 2.4 | 9.21E+15 | 5.40E+11 |
| UGC06917 | 6.17E+09 | 108.7 | 12.7 | 3.48E+10 | 2.06E+14 | 3.64E+13 | 5.6 | 4.41E+14 | 7.46E+10 |
| UGC12632 | 2.95E+09 | 71.7 | 12.6 | 1.51E+10 | 3.87E+13 | 7.59E+12 | 5.1 | 9.18E+13 | 3.57E+10 |
| NGC7814 | 3.89E+10 | 218.9 | 12.2 | 1.35E+11 | 3.24E+15 | 9.32E+14 | 3.5 | 1.13E+16 | 4.71E+11 |
| UGC07524 | 3.55E+09 | 79.5 | 12.1 | 1.78E+10 | 5.62E+13 | 1.12E+13 | 5.0 | 1.36E+14 | 4.29E+10 |
| UGC00731 | 2.57E+09 | 73.3 | 11.6 | 1.45E+10 | 3.88E+13 | 6.91E+12 | 5.6 | 8.36E+13 | 3.11E+10 |
| NGC5585 | 3.72E+09 | 90.3 | 10.9 | 2.07E+10 | 8.44E+13 | 1.51E+13 | 5.6 | 1.83E+14 | 4.50E+10 |
| DDO161 | 2.09E+09 | 66.3 | 10.7 | 1.09E+10 | 2.40E+13 | 4.59E+12 | 5.2 | 5.56E+13 | 2.53E+10 |
| NGC5005 | 9.12E+10 | 262.2 | 10.4 | 1.66E+11 | 5.72E+15 | 3.13E+15 | 1.8 | 3.79E+16 | 1.10E+12 |
| UGC06446 | 2.34E+09 | 82.2 | 10.3 | 1.62E+10 | 5.48E+13 | 7.92E+12 | 6.9 | 9.58E+13 | 2.84E+10 |
| NGC3972 | 8.71E+09 | 132.7 | 10.1 | 4.12E+10 | 3.62E+14 | 7.67E+13 | 4.7 | 9.28E+14 | 1.05E+11 |
| ESO116-G012 | 3.55E+09 | 109.1 | 9.6 | 2.65E+10 | 1.58E+14 | 2.11E+13 | 7.5 | 2.56E+14 | 4.29E+10 |
| UGC10310 | 2.45E+09 | 71.4 | 9.6 | 1.14E+10 | 2.90E+13 | 6.26E+12 | 4.6 | 7.57E+13 | 2.97E+10 |
| UGC11914 | 7.59E+10 | 288.1 | 9.3 | 1.79E+11 | 7.42E+15 | 3.15E+15 | 2.4 | 3.81E+16 | 9.18E+11 |
| UGC04278 | 2.14E+09 | 91.4 | 8.9 | 1.73E+10 | 7.22E+13 | 8.93E+12 | 8.1 | 1.08E+14 | 2.59E+10 |
| UGC06399 | 2.04E+09 | 85.0 | 8.8 | 1.48E+10 | 5.34E+13 | 7.38E+12 | 7.2 | 8.92E+13 | 2.47E+10 |
| UGC04499 | 2.24E+09 | 72.8 | 8.7 | 1.07E+10 | 2.83E+13 | 5.93E+12 | 4.8 | 7.18E+13 | 2.71E+10 |
| UGC06667 | 1.78E+09 | 83.8 | 8.6 | 1.40E+10 | 4.93E+13 | 6.24E+12 | 7.9 | 7.56E+13 | 2.15E+10 |
| UGC08286 | 1.48E+09 | 82.4 | 8.1 | 1.27E+10 | 4.33E+13 | 5.02E+12 | 8.6 | 6.08E+13 | 1.79E+10 |
| UGC07399 | 1.58E+09 | 103.0 | 7.9 | 1.94E+10 | 1.03E+14 | 8.41E+12 | 12.2 | 1.02E+14 | 1.92E+10 |
| UGC08490 | 1.48E+09 | 78.6 | 7.8 | 1.12E+10 | 3.46E+13 | 4.57E+12 | 7.6 | 5.53E+13 | 1.79E+10 |
| NGC0024 | 2.82E+09 | 106.3 | 7.3 | 1.92E+10 | 1.08E+14 | 1.59E+13 | 6.8 | 1.93E+14 | 3.41E+10 |
| UGC05721 | 1.02E+09 | 79.7 | 6.7 | 9.96E+09 | 3.16E+13 | 3.25E+12 | 9.7 | 3.93E+13 | 1.24E+10 |
| UGC04325 | 1.91E+09 | 90.9 | 6.6 | 1.27E+10 | 5.25E+13 | 7.87E+12 | 6.7 | 9.53E+13 | 2.31E+10 |
| UGC07151 | 1.95E+09 | 73.5 | 6.4 | 8.03E+09 | 2.17E+13 | 5.27E+12 | 4.1 | 6.37E+13 | 2.36E+10 |
| NGC3109 | 7.24E+08 | 66.2 | 6.0 | 6.12E+09 | 1.34E+13 | 1.59E+12 | 8.4 | 1.92E+13 | 8.77E+09 |
| UGC08550 | 5.25E+08 | 56.9 | 5.6 | 4.21E+09 | 6.81E+12 | 8.50E+11 | 8.0 | 1.03E+13 | 6.35E+09 |
| UGC01281 | 5.62E+08 | 55.2 | 5.3 | 3.73E+09 | 5.68E+12 | 8.57E+11 | 6.6 | 1.04E+13 | 6.80E+09 |
| UGC07603 | 5.37E+08 | 61.6 | 4.4 | 3.86E+09 | 7.32E+12 | 1.02E+12 | 7.2 | 1.23E+13 | 6.50E+09 |
| UGCA442 | 4.17E+08 | 56.4 | 4.4 | 3.23E+09 | 5.14E+12 | 6.63E+11 | 7.8 | 8.02E+12 | 5.04E+09 |
| NGC3741 | 2.57E+08 | 50.1 | 4.2 | 2.45E+09 | 3.08E+12 | 3.23E+11 | 9.5 | 3.90E+12 | 3.11E+09 |
| DDO064 | 3.63E+08 | 46.1 | 3.5 | 1.72E+09 | 1.83E+12 | 3.86E+11 | 4.8 | 4.67E+12 | 4.39E+09 |
| | | | Mean | 2.01E+11 | 6.11E+15 | 1.46E+15 | 5.4 | 1.76E+16 | 6.03E+11 |
| | | | St. Dev. | 2.95E+11 | 1.29E+16 | 2.67E+15 | 2.4 | 3.23E+16 | 8.12E+11 |
| | | | Median | 7.23E+10 | 5.58E+14 | 8.03E+13 | 5.0 | 9.72E+14 | 1.36E+11 |

*Appendix C: HIFLUGCS Galaxy Cluster Data, Parameters, and Working Equations*

HIghest X-ray FLUx Galaxy Cluster Sample (HIFLUGCS): Table 2 (Tian, 2021)

| ID | Model Input | | | Mass and Energy Parameters | | | | | |
|---|---|---|---|---|---|---|---|---|---|
| Cluster Name | Baryon Mass | Velocity Dispersion | Last Radius | Dynamic Mass | Dynamic Energy | Baryon Energy | 'Mass Discrepancy' | Virial Energy | Total Mass |
| n=29 | $M_{Bar}$ | $\sigma_{last}$ | $r_{last}$ | $M_{Dyn}=R_l\sigma_l^2/G$ | $E_{Dyn}=R_l\sigma_l^4/2G$ | $E_{Bar}=M_{Bar}\sigma_l^2/2$ | $D=M_{Dyn}/M_{Bar}$ $D=E_{Dyn}/E_{Bar}$ | $E_{Vir}=12.1E_{Bar}$ | $M_{Vir}=12.1M_{Bar}$ |
| [ID] | [$M_\odot$] | [kms$^{-1}$] | [kpc] | [$M_\odot$] | [$M_\odot$km$^2$s$^{-2}$] | [$M_\odot$km$^2$s$^{-2}$] | [Ratio] | [$M_\odot$km$^2$s$^{-2}$] | [$M_\odot$] |
| NGC 4636 | 2.39E+11 | 229 | 29 | 3.54E+11 | 9.27E+15 | 6.28E+15 | 1.5 | 7.59E+16 | 2.90E+12 |
| Fornax | 6.71E+11 | 252 | 1384 | 2.04E+13 | 6.49E+17 | 2.13E+16 | 30.4 | 2.58E+17 | 8.12E+12 |
| A3526 | 1.19E+13 | 890 | 779 | 1.43E+14 | 5.68E+19 | 4.70E+18 | 12.1 | 5.68E+19 | 1.43E+14 |
| A1060 | 9.42E+12 | 634 | 389 | 3.64E+13 | 7.31E+18 | 1.89E+18 | 3.9 | 2.29E+19 | 1.14E+14 |
| A262 | 1.17E+13 | 551 | 2125 | 1.50E+14 | 2.28E+19 | 1.78E+18 | 12.8 | 2.16E+19 | 1.42E+14 |
| A3581 | 5.70E+12 | 436 | 649 | 2.87E+13 | 2.73E+18 | 5.42E+17 | 5.0 | 6.56E+18 | 6.90E+13 |
| A4038 | 1.92E+13 | 773 | 513 | 7.13E+13 | 2.13E+19 | 5.75E+18 | 3.7 | 6.95E+19 | 2.33E+14 |
| A2634 | 1.49E+13 | 731 | 1220 | 1.52E+14 | 4.05E+19 | 3.99E+18 | 10.2 | 4.83E+19 | 1.81E+14 |
| A496 | 2.96E+13 | 648 | 805 | 7.86E+13 | 1.65E+19 | 6.22E+18 | 2.7 | 7.53E+19 | 3.59E+14 |
| A2063 | 2.14E+13 | 779 | 844 | 1.19E+14 | 3.61E+19 | 6.49E+18 | 5.6 | 7.85E+19 | 2.59E+14 |
| A2052 | 2.09E+13 | 475 | 622 | 3.26E+13 | 3.68E+18 | 2.36E+18 | 1.6 | 2.85E+19 | 2.53E+14 |
| A2147 | 4.12E+13 | 811 | 1146 | 1.75E+14 | 5.76E+19 | 1.36E+19 | 4.3 | 1.64E+20 | 4.99E+14 |
| A576 | 2.14E+13 | 923 | 1582 | 3.13E+14 | 1.34E+20 | 9.11E+18 | 14.7 | 1.10E+20 | 2.59E+14 |
| A3571 | 5.42E+13 | 841 | 999 | 1.64E+14 | 5.81E+19 | 1.92E+19 | 3.0 | 2.32E+20 | 6.56E+14 |
| A2589 | 1.90E+13 | 610 | 1148 | 9.93E+13 | 1.85E+19 | 3.54E+18 | 5.2 | 4.28E+19 | 2.30E+14 |
| A2657 | 1.77E+13 | 789 | 666 | 9.64E+13 | 3.00E+19 | 5.50E+18 | 5.5 | 6.65E+19 | 2.14E+14 |
| A119 | 4.49E+13 | 648 | 1240 | 1.21E+14 | 2.54E+19 | 9.42E+18 | 2.7 | 1.14E+20 | 5.43E+14 |
| A3558 | 6.68E+13 | 820 | 1558 | 2.44E+14 | 8.19E+19 | 2.25E+19 | 3.6 | 2.72E+20 | 8.09E+14 |
| A1644 | 4.46E+13 | 901 | 1060 | 2.00E+14 | 8.12E+19 | 1.81E+19 | 4.5 | 2.19E+20 | 5.39E+14 |
| A3562 | 2.54E+13 | 729 | 1269 | 1.57E+14 | 4.17E+19 | 6.75E+18 | 6.2 | 8.17E+19 | 3.07E+14 |
| A4059 | 2.39E+13 | 666 | 926 | 9.55E+13 | 2.12E+19 | 5.30E+18 | 4.0 | 6.41E+19 | 2.89E+14 |
| A3391 | 3.35E+13 | 885 | 1815 | 3.31E+14 | 1.29E+20 | 1.31E+19 | 9.9 | 1.59E+20 | 4.05E+14 |
| A85 | 6.97E+13 | 934 | 1536 | 3.12E+14 | 1.36E+20 | 3.04E+19 | 4.5 | 3.68E+20 | 8.43E+14 |
| A133 | 2.47E+13 | 803 | 1021 | 1.53E+14 | 4.94E+19 | 7.95E+18 | 6.2 | 9.62E+19 | 2.98E+14 |
| A3158 | 3.96E+13 | 985 | 1581 | 3.57E+14 | 1.73E+20 | 1.92E+19 | 9.0 | 2.33E+20 | 4.79E+14 |
| A3266 | 8.75E+13 | 1226 | 2848 | 9.96E+14 | 7.48E+20 | 6.58E+19 | 11.4 | 7.96E+20 | 1.06E+15 |
| A1795 | 5.20E+13 | 831 | 3087 | 4.96E+14 | 1.71E+20 | 1.80E+19 | 9.5 | 2.17E+20 | 6.29E+14 |
| A2029 | 8.59E+13 | 844 | 2751 | 4.56E+14 | 1.62E+20 | 3.06E+19 | 5.3 | 3.70E+20 | 1.04E+15 |
| A2142 | 1.39E+14 | 1062 | 3012 | 7.90E+14 | 4.46E+20 | 7.82E+19 | 5.7 | 9.46E+20 | 1.68E+15 |
| | | | Mean | 2.2E+14 | 9.56E+19 | 1.41E+19 | 7.0 | 1.71E+20 | 4.32E+14 |
| | | | St. Dev. | 2.3E+14 | 1.54E+20 | 1.82E+19 | 5.7 | 2.20E+20 | 3.71E+14 |
| | | | Median | 1.5E+14 | 4.17E+19 | 6.75E+18 | 5.3 | 8.17E+19 | 2.98E+14 |